\documentclass[11pt]{article}

\usepackage{amssymb}
\usepackage{amsmath}

\newtheorem{theorem}{Theorem}
\newtheorem{proposition}{Proposition}
\newtheorem{lemma}{Lemma}

\newenvironment{proof}{\mbox{\sc Proof:}}{$\Box$}
\newenvironment{proof*}{\mbox{\sc Proof:}\hspace{0.3 em}{\bf (*)}\hspace{0.3 em}}{$\Box$}

\renewcommand{\Re}{\mathbb{R}}

\begin{document}

\title{Instantaneous Arbitrage and the CAPM}
\author{Lars Tyge Nielsen \\ Department of Mathematics \\ Columbia University}
\date{January 2019\thanks{Previous Version May 2006}}
\maketitle

\begin{abstract}

This paper studies the concept of instantaneous arbitrage in continuous time and its relation to the instantaneous CAPM.
Absence of instantaneous arbitrage
is equivalent to the existence of a trading strategy which satisfies the CAPM beta pricing relation in place of the market.
Thus the difference between the arbitrage argument and the CAPM argument in
Black and Scholes \cite[1973]{Black-Scholes:73} is this: the arbitrage argument assumes that there exists some portfolio satisfying the capm equation, whereas the CAPM argument assumes, in addition, that this portfolio is the market portfolio.

\end{abstract}

\section{Introduction}

This paper studies the concept of instantaneous arbitrage in continuous time and its relation to the instantaneous CAPM.

An instantaneous arbitrage trading strategy is an instantaneously riskless trading strategy whose instantaneous expected excess return is always non-negative and sometimes positive.

We define a market to be instantaneously arbitrage-free if it is not possible to construct a zero-value instantaneous arbitrage trading strategy. This definition is independent of any choice of a potentially non-unique interest rate process, because for trading strategies with zero value (zero cost), the excess return equals the return and does not involve the interest rate.

If the market is free of instantaneous arbitrage according to this definition, then it follows that any interest rate process is unique in the sense that two money market accounts must have associated interest rate processes that are almost everywhere identical.

Once a particular interest rate process has been fixed,
absence of instantaneous arbitrage can be defined in various equivalent ways.
It is equivalent to the non-existence of an arbitrage trading strategy (zero-value or not), and to the non-existence of a self-financing arbitrage trading strategy.
It is also equivalent to the condition that every instantaneously riskless trading strategy has zero expected excess rate of return almost everywhere. 

Apart from the various ways it can be defined, 
absence of instantaneous arbitrage is equivalent to the existence of a vector of prices of risk, and it is equivalent to the existence of a trading strategy whose dispersion is a vector of prices of risk.
The dispersion of such a trading strategy is in fact a minimal vector of prices of risk, in the sense that its Euclidean length is less than that of any other vector of prices of risk.

Most importantly,
absence of instantaneous arbitrage
is equivalent to the existence of a trading strategy which satisfies the CAPM beta pricing relation in place of the market.

According to the CAPM, in equilibrium, the expected excess return to any asset equals its beta with respect to the market portfolio times the expected excess return on the market portfolio.  
It is well known that this CAPM relation may be satisfied by portfolios other than the market portfolio, even if the market is not in equilibrium.
Indeed, Roll \cite[1977]{Roll:77b} showed that irrespective of the equilibrium assumption, a portfolio satisfies the CAPM relation if and only if it is on the mean-variance frontier.

In continuous time, it is equally true that the CAPM relation may be satisfied by some portfolio even if it is not satisfied by the market and even if the market is not in equlibrium. What is required is that the market should be free of instantaneous arbitrage opportunities. The main result of this paper, which is stated in Theorem~\ref{arfbeta-t}, says that there exists a trading strategy that satisfies the CAPM relation if and only if the market is instaneously arbitrage-free.

This throws light on the two competing approaches used by 
Black and Scholes \cite[1973]{Black-Scholes:73} in deriving their
derive their partial differential equation: absence of instantaneous arbitrage, and the CAPM.

Black and Scholes \cite[1973]{Black-Scholes:73} attributed the instantaneous-arbitrage argument to Robert C. Merton. See also Black \cite[1989]{Black:89:How}.
Merton used the argument in \cite[1973]{Merton:73:option}. 
Duffie \cite[1998]{Duffie:98:Black} wrote that it
``truly revolutionized modern finance theory'', and
Schaefer \cite[1998]{Schaefer:98} described it as a ``seminal'' and ``critical'' observation.

Our result implies that although the assumption of absence of instantaneous arbitrage is somewhat weaker than the statement that the market satisfies the CAPM relation, it is not fundamentally different, because it is exactly equivalent to the statement that some trading strategy satisfies the CAPM relation. The latter is clearly sufficient in the derivation of the PDE from the CAPM in Black and Scholes \cite[1973]{Black-Scholes:73}.

Recall that none of these assumptions is sufficient for deriving the Black--Scholes formula.
See the discussion in Nielsen \cite[1999, Section~~6.12]{Nielsen:99}. At best, they imply the Black--Scholes PDE. However, as Black \cite[1976]{Black:76} acknowledged, the PDE does not have a unique solution.
Therefore, the PDE alone does not imply that the option price must be given by the Black--Scholes formula.

The paper is organized as follows. Section~\ref{model-sect} describes the model. It is a continuous-time trading model with It\^{o} gains processes and general dividend processes, as in Nielsen \cite[2007]{Nielsen:07}.
Section~\ref{def-sect} defines instantaneous arbitrage in various equivalent ways and observes that the interest rate process is unique in an instantaneously arbitrage-free market.
Section~\ref{CAPM-sect} shows that absence of instantaneous arbitrage
is equivalent to the existence of a vector of prices of risk, to the existence of a trading strategy whose dispersion is a minimal vector of prices of risk, and to the existence of a trading strategy which satisfies the CAPM relation.

The proofs are in Appendix~\ref{proofs-app}. They involve a combination of linear algebra and measure theory. Because we are dealing with stochastic processes, we need to know that solutions of linear equations, when they exist but are not unique, can be chosen so as to be measurable functions of the vectors and matrices that define the equations. These things are spelled out in Appendix~\ref{linalg-app}. In particular, the proof that absence of instantaneous arbitrage
implies the existence of a trading strategy whose dispersion is vector of prices of risk relies on a dual characterization of the existence of a measurable and adapted solution to a linear equation whose parameters are measurable and adapted processes.

\section{Securities and Trading Strategies}

\label{model-sect}

We consider a securities market where the uncertainty is represented by
a complete probability space
$(\Omega,{\cal F},P)$ with a filtration
$F = \{ {\cal F}_{t} \}_{t \in {\cal T}}$
and a $K$-dimensional process $W$, which
is a Wiener process relative to $F$.

A \emph{cumulative dividend process} is a measurable adapted process $D$
with $D(0) = 0$.

Suppose a security has cumulative dividend process $D$ and \emph{price process} $S$.
Define the \emph{cumulative gains process} $G$ of this security as the sum of the
price process and the cumulative dividend process:
\[ G = S + D \]
Assume that $G$ is an It\^{o} process.
It follows that $G$ will be continuous, adapted, and measurable.  Since $D$ is adapted and
measurable, so is $S$.  Since $D(0) = 0$, $G(0) = S(0)$.

An $(N+1)$-dimensional \emph{securities market model} (based on $F$ and $W$)
will be a pair $(\bar{S},\bar{D})$ of measurable and adapted processes $\bar{S}$ and $\bar{D}$ of dimension $N+1$, interpreted as a vector of price processes and a vector of cumulative dividends processes, such that $\bar{D}(0) = 0$ and such that $\bar{G} = \bar{S}+\bar{D}$ is 
an It\^{o} process with respect to $F$ and $W$.
The process $\bar{G} = \bar{S}+\bar{G}$
is the cumulative gains processes corresponding to $(\bar{S},\bar{D})$.

Write
\[ \bar{G}(t) = \bar{G}(0) + \int_{0}^{t}\bar{\mu} \, ds + 
 \int_{0}^{t} \bar{\sigma} \, dW \]
where $\bar{\mu}$ is an $N+1$ dimensional vector process
in $\mathcal{L}^{1}$ and $\bar{\sigma}$ is an
$(N+1) \times K$ dimensional matrix valued process in $\mathcal{L}^{2}$.

Here, $\mathcal{L}^{1}$ is the set of adapted, measurable, and pathwise integrable processes,
and $\mathcal{L}^{2}$
is the set of adapted, measurable, and pathwise square integrable processes.

A \emph{trading strategy} is an adapted, measurable $(N+1)$-dimensional row-vector-valued process $\bar{\Delta}$.

The \emph{value process} of a trading strategy $\bar{\Delta}$ 
in securities model $(\bar{S},\bar{D})$ is the process
$\bar{\Delta}\bar{S}$.

The set of trading strategies $\bar{\Delta}$
such that $\bar{\Delta} \bar{\mu} \in \mathcal{L}^{1}$
and $\bar{\Delta} \bar{\sigma} \in \mathcal{L}^{2}$,
will be denoted $\mathcal{L}(\bar{G})$.

In general, if $X$ is an $n$-dimensional It\^{o} process,
\[ X(t) = X(0) + \int_{0}^{t} a \, ds + \int_{0}^{t} b \,dW \]
then $\mathcal{L}(X)$ is the set of adapted, measurable, $(n \times K)$-dimensional 
processes $\gamma$ such that
$\gamma a \in \mathcal{L}^{1}$ 
and $\gamma b \in \mathcal{L}^{2}$.

If $\bar{\Delta}$ is a trading strategy in $\mathcal{L}(\bar{G})$,
then the \emph{cumulative gains process} of $\bar{\Delta}$, measured relative to the securities market model $(\bar{S},\bar{D})$, is the process
$\mathcal{G}(\bar{\Delta};\bar{G})$ defined by
\[ \mathcal{G}(\bar{\Delta};\bar{G})(t) =
\bar{\Delta}(0)\bar{G}(0) +
 \int_{0}^{t} \bar{\Delta}\, d\bar{G} \]
for all $t \in\mathcal{T}$.

A trading strategy $\bar{\Delta}$
in $\mathcal{L}(\bar{G})$
is \emph{self-financing} with respect to $(\bar{S},\bar{D})$ if
\[ \bar{\Delta}\bar{S} = \mathcal{G}(\bar{\Delta};\bar{G}) \]
or
\[ \bar{\Delta}(t)\bar{S}(t) = 
\bar{\Delta}(0)\bar{S}(0) +
 \int_{0}^{t} \bar{\Delta} \, d\bar{G} \]

Generally, if $\bar{\Delta}$ is a trading strategy in $\mathcal{L}(\bar{G})$ which may not be self-financing, then the
\emph{cumulative dividend process} of $\bar{\Delta}$  
with respect to $(\bar{S},\bar{D})$ is the process
$\mathcal{D}(\bar{\Delta};\bar{S},\bar{D})$ defined by
\[ \bar{\Delta}\bar{S} +\mathcal{D}(\bar{\Delta};\bar{S},\bar{D}) =
\mathcal{G}(\bar{\Delta};\bar{G}) \]
The process
$\mathcal{D}(\bar{\Delta};\bar{S},\bar{D})$ is adapted and measurable
and has initial value $\mathcal{D}(\bar{\Delta};\bar{S},\bar{D})(0) = 0$.

A \emph{money market account} for $(\bar{S},\bar{D})$
is a self-financing trading strategy $\bar{b}$ (or a security that
pays no dividends) whose value process is positive and instantaneously riskless (has zero dispersion).
We denote its value process by $M$: $M = \bar{b}\bar{S}$.

If $M$ is the value process of a money market account, then it must have the form
\[ M(t) = M(0)\exp\left\{ \int_{0}^{t} r \, ds \right\} \]
for some $r \in\mathcal{L}^{1}$ (the \emph{interest rate} process) and some $M(0) > 0$.

In changing to units of the money market account, we rely on the formulas developed in Nielsen \cite[2007]{Nielsen:07}.

Let $D$ be a cumulative dividend process.
Then $Dr \in \mathcal{L}^{1}$ if and only if
$D \in \mathcal{L}(1/M) = \mathcal{L}(M)$.
If so, then the cumulative dividend process in units of the money market account is
\[ D^{1/M}(t) = D(t)/M(t) + \int_{0}^{t} D\frac{r}{M} \, d s \]

If $(S,D)$ is a security model with $D \in \mathcal{L}(M)$, and if the associated gains process $G = S+D$ has differential
\[ dG = \mu \, dt + \sigma \, dW \]
then the cumulative gains process in units of the money market account is
\[ G^{1/M}(t) = 
G(t)/M(t) 
+ \int_{0}^{t} D\frac{r}{M} \, d s \]
and
\[ dG^{1/M}(t) = 
\frac{\mu - rS}{M} \, dt
+ \frac{\sigma}{M} \, dW \]

\section{Definitions of Arbitrage-Free Markets}

\label{def-sect}

To begin with, we define instantaneous arbitrage in a way that does not assume the existence of a money market account or an instantaneous interest rate process. On this basis we show that absence of instantaneous arbitrage implies uniqueness of the instantaneous interest rate process and of the value process of the money market account (up to a positive scaling factor). We then re-formulate instantaneous arbitrage in various equivalent ways which do involve the instantaneous interest rate process.

If $\bar{\Delta}$ is a trading strategy (not necessarily self-financing), then the processes $\bar{\Delta}\bar{\sigma}\bar{\sigma}^{\top}\bar{\Delta}^{\top}$
and $\sqrt{\bar{\Delta}\bar{\sigma}\bar{\sigma}^{\top}\bar{\Delta}^{\top}}$
will be called the \emph{instantaneous dollar return variance},
and
the \emph{instantaneous dollar return standard deviation}
of $\bar{\Delta}$,
respectively.

A trading strategy $\bar{\Delta}$ is \emph{instantaneously riskless} if $\bar{\Delta}\bar{\sigma} = 0$ almost everywhere.

A \emph{zero-value instantaneous arbitrage trading strategy} is an instantaneously riskless trading strategy $\bar{\Delta}$ such that $\bar{\Delta}\bar{S} = 0$ almost everywhere, $\bar{\Delta}\bar{\mu} \geq 0$ almost everywhere, and $\bar{\Delta}\bar{\mu} > 0$ on a set of positive measure.

A zero-value instantaneous arbitrage trading strategy is not supposed to be self-financing. Since its value is always zero, it will normally be paying dividends all the time.

Say that a securities
market model $(\bar{S},\bar{D})$ is \emph{instantaneously arbitrage-free} if there exists no zero-value instantaneous arbitrage trading strategy.

\begin{proposition}
\label{mmaun-p}
Suppose the securities market model $(\bar{S},\bar{D})$ is instantaneously arbitrage-free.
If $\bar{b}_{1}$ and $\bar{b}_{2}$ are money market accounts with interest rate processes $r_{1}$ and $r_{2}$ and value processes $M_{1}$ and $M_{2}$, then $r_{1}$ and $r_{2}$ are almost everywhere identical, and $M_{1}/M_{1}(0)$ and $M_{2}/M_{2}(0)$ are indistinguishable.
\end{proposition}

The proofs of Propostion~\ref{mmaun-p} and other results in this and the following section are in Appendix~\ref{proofs-app}.

Now let $\bar{b}$ be a money market account with value process $M$ and interest rate process $r$.

If $\bar{\Delta}$ is a trading strategy (not necessarily self-financing), then the processes $\bar{\Delta}(\bar{\mu}-r\bar{S})$
will be called the \emph{instantaneous excess expected dollar return}
of $\bar{\Delta}$.

Assume that $\bar{D} \in \mathcal{L}(1/M)$.

An \emph{instantaneous arbitrage trading strategy} is an instantaneously riskless trading strategy $\bar{\Delta} \in \mathcal{L}(\bar{G}^{1/M})$ such that
$\bar{\Delta}(\bar{\mu}-r\bar{S}) \geq 0$
almost everywhere, and
$\bar{\Delta}(\bar{\mu}-r\bar{S}) > 0$
on a set of positive measure.

Recall
that
\[ d\bar{G}^{1/M} =
\frac{1}{M}(\bar{\mu} - r\bar{S}) \, dt + \frac{1}{M} \bar{\sigma} \, dW \]
It follows that
a zero-value trading strategy is in $\mathcal{L}(\bar{G})$ if and only if it is in $\mathcal{L}(\bar{G}^{1/M})$.
Therefore, the definition of an instantaneous arbitrage trading strategy is
consistent with the earlier definition of a zero-value instantaneous arbitrage trading strategy.  A
zero-value instantaneous arbitrage trading strategy is nothing other than an instantaneous arbitrage trading strategy $\bar{\Delta}$ such that $\bar{\Delta}\bar{S} = 0$ almost everywhere.

The following proposition shows that a number of possible definitions of the concept of an instantaneously arbitrage-free securities market model are equivalent.

\begin{proposition}
\label{exinstarbfree-p}
The following statements are equivalent:
\begin{enumerate}
\item
$(\bar{S},\bar{D})$ is instantaneously arbitrage-free (there exists no zero-value instantaneous arbitrage trading strategy)
\item
There exists no instantaneous arbitrage trading strategy
\item
There exists no self-financing instantaneous arbitrage trading strategy
\item
Every instantaneously riskless trading strategy has zero instantaneous expected excess return almost everywhere
\end{enumerate}
\end{proposition}

\section{Prices of Risk and the CAPM}

\label{CAPM-sect}

A vector of
instantaneous \emph{prices of risk} is an adapted measurable $K$-dimensional row-vector-valued process $\lambda$ such that
\[ \bar{\mu} - r \bar{S} = \bar{\sigma} \lambda^{\top} \]
almost everywhere\footnote{Some authors, including Nielsen \cite[1999]{Nielsen:99}, require in addition that $\lambda \in \mathcal{L}^{2}$.}.

Proposition~\ref{noinstarbt-p} states the main characteristics of an instantaneously arbitrage-free securities market model.

\begin{proposition}
\label{noinstarbt-p}
The following statements are equivalent:
\begin{enumerate}
\item
$(\bar{S},\bar{D})$ is instantaneously arbitrage free
\item
There exists a vector of instantaneous prices of risk
\item
There exists a trading strategy $\bar{\psi}$ such that
\[ \bar{\mu} - r \bar{S} = \bar{\sigma}\bar{\sigma}^{\top}\bar{\psi}^{\top} \]
almost everywhere
\end{enumerate}
\end{proposition}

If $\bar{\psi}$ is a trading strategy such as the one in (3) of Proposition~\ref{noinstarbt-p}, then according to Proposition~\ref{lambdastar-p} below,
the process $\lambda^{*} = \bar{\psi}\bar{\sigma}$ will be the minimal vector of prices of risk, in the sense that for any other vector $\lambda$ of prices of risk, $\lambda^{*}\lambda^{*\top} \leq \lambda\lambda^{\top}$ almost everywhere.

\begin{proposition}
\label{lambdastar-p}
Suppose $\lambda$ is a vector of instantaneous prices of risk and $\bar{\psi}$ is a trading strategy such that
\[ \bar{\mu} - r \bar{S} = \bar{\sigma}\bar{\sigma}^{\top}\bar{\psi}^{\top} \]
almost everywhere. Set
$\lambda^{*} = \bar{\psi} \bar{\sigma}$.
Then $\lambda^{*}\lambda^{*\top} \leq \lambda\lambda^{\top}$
almost everywhere.
\end{proposition}

If $\bar{\Delta}$ is a trading strategy, let $b_{\bar{\Delta}}$ be the vector of \emph{fundamental betas} of the individual securities with respect to $\bar{\Delta}$:
\[ b_{\bar{\Delta}} = \left\{ \begin{array}{cc} \frac{1}{\bar{\Delta}\bar{\sigma}\bar{\sigma}^{\top}\bar{\Delta}^{\top}}
\bar{\sigma}\bar{\sigma}^{\top}\bar{\Delta}^{\top}
& \mbox{ if } \bar{\Delta} \bar{\sigma}\bar{\sigma}^{\top}\bar{\Delta}^{\top} \neq 0 \\
0 & \mbox{ otherwise} \end{array} \right. \]

Say that a trading strategy $\bar{\Delta}$ \emph{satisfies the CAPM equation} if
\[ \bar{\mu} - r\bar{S} = b_{\bar{\Delta}}\bar{\Delta}(\bar{\mu} - r\bar{S}) \]
almost everywhere.

\begin{theorem}
\label{arfbeta-t}
$(\bar{S},\bar{D})$ is instantaneously arbitrage free if and only if
there exists a trading strategy $\bar{\Delta}$ which satisfies the CAPM equation.
\end{theorem}

Theorem~\ref{arfbeta-t} implies
that the difference between the arbitrage argument and the CAPM argument in
Black and Scholes \cite[1973]{Black-Scholes:73} is this: the arbitrage argument assumes that there exists some portfolio satisfying the capm equation, whereas the CAPM argument assumes, in addition, that this portfolio is the market portfolio.

\appendix

\section{Appendix A: Proofs}

\label{proofs-app}

Most proofs are in this appendix. The proof of Proposition~\ref{noinstarbt-p} depends on
some concepts of ``measurable linear algebra,'' which are developed in Appendix~\ref{linalg-app}, culminating in 
Proposition~\ref{conj-p}.

{\sc Proof of Proposition~\ref{mmaun-p}}:

Suppose it is not true that $r_{1}$ and $r_{2}$ are almost everywhere identical.  Assume without loss of generality that $r_{2} > r_{1}$ on a set of positive measure.

Define a trading strategy $\bar{b}$ by
\[ \bar{b} =
1_{r_{1} \geq r_{2}}\left(\frac{1}{M_{1}}\bar{b}_{1}-\frac{1}{M_{2}}\bar{b}_{2}\right)
+ 1_{r_{2} > r_{1}}\left(\frac{1}{M_{2}}\bar{b}_{2}-\frac{1}{M_{1}}\bar{b}_{1}\right) \]
Then
\begin{eqnarray*}
\bar{b}\bar{S}
& = &
1_{r_{1} \geq r_{2}}\left(\frac{1}{M_{1}}\bar{b}_{1}\bar{S}-\frac{1}{M_{2}}\bar{b}_{2}\bar{S}\right)
+ 1_{r_{2} > r_{1}}\left(\frac{1}{M_{2}}\bar{b}_{2}\bar{S}-\frac{1}{M_{1}}\bar{b}_{1}\bar{S}\right) \\
& = &
1_{r_{1} \geq r_{2}}\left(\frac{1}{M_{1}}M_{1}-\frac{1}{M_{2}}M_{2}\right)
+ 1_{r_{2} > r_{1}}\left(\frac{1}{M_{2}}M_{2}-\frac{1}{M_{1}}M_{1}\right) \\
& = &
0
\end{eqnarray*}
\begin{eqnarray*}
\bar{b}\bar{\mu}
& = &
1_{r_{1} \geq r_{2}}\left(\frac{1}{M_{1}}\bar{b}_{1}\bar{\mu}-\frac{1}{M_{2}}\bar{b}_{2}\bar{\mu}\right)
+ 1_{r_{2} > r_{1}}\left(\frac{1}{M_{2}}\bar{b}_{2}\bar{\mu}-\frac{1}{M_{1}}\bar{b}_{1}\bar{\mu}\right) \\
& = &
1_{r_{1} \geq r_{2}}\left(\frac{1}{M_{1}}r_{1}M_{1}-\frac{1}{M_{2}}r_{2}M_{2}\right)
+ 1_{r_{2} > r_{1}}\left(\frac{1}{M_{2}}r_{2}M_{2}-\frac{1}{M_{1}}r_{1}M_{1}\right) \\
& = &
1_{r_{1} \geq r_{2}}\left(r_{1}-r_{2}\right)
+ 1_{r_{2} > r_{1}}\left(r_{2}-r_{1}\right)
\end{eqnarray*}
which is non-negative almost everywhere and positive on a set of positive measure,
and
\[ \bar{b}\bar{\sigma} =
1_{r_{1} \geq r_{2}}\left(\frac{1}{M_{1}}\bar{b}_{1}\bar{\sigma}-\frac{1}{M_{2}}\bar{b}_{2}\bar{\sigma}\right)
+ 1_{r_{2} > r_{1}}\left(\frac{1}{M_{2}}\bar{b}_{2}\bar{\sigma}-\frac{1}{M_{1}}\bar{b}_{1}\bar{\sigma}\right) = 0 \]
almost everywhere.
Hence, $\bar{\Delta}$ is a zero-value instantaneous arbitrage trading strategy, a contradiction.

It follows that $r_{1}$ and $r_{2}$ are almost everywhere identical.
This implies that $M_{1}/M_{1}(0)$ and $M_{2}/M_{2}(0)$ are indistinguishable.

$\Box$

\begin{lemma}
\label{selfarb2-l}
Let $\bar{b}$ be a money market account with value process $M$.
Assume that $\bar{D} \in \mathcal{L}(1/M)$.
Let $\bar{\Delta}$ be a trading strategy.  
The trading strategy
\[ \bar{\Theta} = \bar{\Delta} + \mathcal{D}\left( \bar{\Delta}; \bar{S}/M,\bar{D}^{1/M} \right)\bar{b} \]
is self-financing with
\[ \bar{\Theta}\bar{S}/M = \mathcal{G}\left( \bar{\Delta}; \bar{G}^{1/M} \right) \]
If $\bar{\Delta}$ is an instantaneous arbitrage trading strategy, then so is $\bar{\Theta}$.
\end{lemma}
\begin{proof}
It is easily seen that
\[ \mathcal{D}\left(\bar{\Theta};\bar{S}/M,\bar{D}^{1/M}\right) = \mathcal{D}\left( \bar{\Delta}; \bar{S}/M,\bar{D}^{1/M} \right) - \mathcal{D}\left( \bar{\Delta}; \bar{S}/M,\bar{D}^{1/M} \right) = 0 \]
which implies that $\bar{\Theta}$ is self-financing.
The process
\[ \mathcal{D}\left( \bar{\Delta}; \bar{S}/M,\bar{D}^{1/M} \right)\bar{b} \]
has zero cumulative gains process with respect to $(\bar{S}/M,\bar{D}^{1/M})$. Hence,
\[ \bar{\Theta}\bar{S}/M = \mathcal{G}\left( \bar{\Theta}; \bar{G}^{1/M} \right) =
 \mathcal{G}\left( \bar{\Delta}; \bar{G}^{1/M} \right) \]
Observe that
\[ \bar{\Theta}(\bar{\mu}-r\bar{S}) = \bar{\Delta}(\bar{\mu}-r\bar{S}) \]
and $\bar{\Theta}\bar{\sigma} = \bar{\Delta}\bar{\sigma}$ almost everywhere.  It follows that if $\bar{\Delta}$ is an instantaneous arbitrage trading strategy, then so is $\bar{\Theta}$.
\end{proof}

\medskip
{\sc Proof of Proposition~\ref{exinstarbfree-p}}:

It is useful to observe that Statement (4) is equivalent to the following:

\begin{itemize}
\item [(5)]
There exists no trading strategy $\bar{\Delta}$ such that
on a set of positive measure,
$\bar{\Delta} \bar{\sigma}\bar{\sigma}^{\top}\bar{\Delta}^{\top} = 0$
and
$\bar{\Delta}(\bar{\mu}-r\bar{S}) > 0$
\end{itemize}

It is obvious that (5) implies Statement (4) in the proposition.
Conversely, to show that (4) implies (5),
suppose there exists a trading strategy $\bar{\Delta}$ as in (5).

Let $\mathcal{A} \subset \Omega \times \mathcal{T}$ be the set of $(\omega,t)$ such that
\[ \bar{\Delta}(\omega,t) (\bar{\mu}(\omega,t)-r(\omega,t)\bar{S}(\omega,t)) > 0 \]
Then the indicator function $1_{\mathcal{A}}$ is a measurable and adapted process, and $\mathcal{A}$ is measurable with positive measure.

Define the process $\bar{\Theta}$ by
$\bar{\Theta} = \bar{\Delta}$ on $\mathcal{A}$ and $\bar{\Theta} = 0$ outside of
$\mathcal{A}$.
Then $\bar{\Theta}$ is measurable and adapted, and, hence, it is a trading strategy.
It is an instantaneously riskless trading strategy with positive instantaneous expected excess return on a set of positive measure, contradicting (4).

(2) equivalent to (3): Follows from Lemma~\ref{selfarb2-l}.

(5) implies (2): Obvious.

(2) implies (1):

If $(\bar{S},\bar{D})$ is not instantaneously arbitrage free, then there exists a zero-value instantaneous arbitrage trading strategy, which is, in particular, an instantaneous arbitrage trading strategy.

(1) implies (5):

Suppose there exists a
trading strategy $\bar{\Delta}$ such that
on a set of positive measure,
$\bar{\Delta} \bar{\sigma}\bar{\sigma}^{\top}\bar{\Delta}^{\top} = 0$
and
$\bar{\Delta}(\bar{\mu}-r\bar{S}) > 0$.

Let $\mathcal{A} \subset \Omega \times \mathcal{T}$ be the set of $(\omega,t)$ such that
\[ \bar{\Delta}(\omega,t)\bar{\sigma}(\omega,t)\bar{\sigma}(\omega,t)^{\top}\bar{\Delta}(\omega,t)^{\top} = 0 \]
and
\[ \bar{\Delta}(\omega,t) (\bar{\mu}(\omega,t)-r(\omega,t)\bar{S}(\omega,t)) > 0 \]
Then the indicator function $1_{\mathcal{A}}$ is a measurable and adapted process, and $\mathcal{A}$ is measurable with positive measure.

Define a process $\bar{\Theta}$ by
\[ \bar{\Theta} = \bar{\Delta} - \frac{\bar{\Delta}\bar{S}}{M}\bar{b} \]
on $\mathcal{A}$ and $\bar{\Theta} = 0$ outside of $\mathcal{A}$.
Then $\bar{\Theta}$ is measurable and adapted, and, hence, it is a trading strategy.

On $\mathcal{A}$,
$\bar{\Theta}\bar{\sigma}\bar{\sigma}^{\top}\bar{\Theta}^{\top} = 0$,
\[ \bar{\Theta}\bar{S} = \bar{\Delta}\bar{S} - \frac{\bar{\Delta}\bar{S}}{M}\bar{b}\bar{S} = 0 \]
and
\[ \bar{\Theta}\bar{\mu} = \bar{\Delta}\bar{\mu} - \frac{\bar{\Delta}\bar{S}}{M}\bar{b}\bar{\mu}
= \bar{\Delta}(\bar{\mu} - r\bar{S}) > 0  \]
Outside of $\mathcal{A}$, $\bar{\Theta} = 0$, and, hence,
$\bar{\Theta}\bar{\sigma}\bar{\sigma}^{\top}\bar{\Theta}^{\top} = 0$,
$\bar{\Theta}\bar{S} = 0$, and
$\bar{\Theta}\bar{\mu} = 0$.
This implies that
$\bar{\Theta}$ is a zero-value instantaneous arbitrage trading strategy.
Hence, $(\bar{S},\bar{D})$ is not instantaneously arbitrage free.

$\Box$

\medskip
{\sc Proof of Proposition~\ref{noinstarbt-p}}:

(3) implies (2):

Set $\lambda = \bar{\sigma} \bar{\psi}$.

(2) implies (1):

If an instantaneous arbitrage trading strategy $\bar{\Delta}$ exists, then
\[ \bar{\Delta} (\bar{\mu} - r \bar{S}) = \bar{\delta}\bar{\sigma} \lambda^{\top} = 0 \]
almost everywhere, a contradiction.

(1) implies (3):

From (5) of the proof of Propostion~\ref{exinstarbfree-p}, we know that there exists no
trading strategy $\bar{\Delta}$ such that
on a set of positive measure,
$\bar{\Delta} \bar{\sigma}\bar{\sigma}^{\top}\bar{\Delta}^{\top} = 0$
and
$\bar{\Delta}(\bar{\mu}-r\bar{S}) = 1$

From Proposition~\ref{conj-p} in Appendix~\ref{linalg-app}, it then follows that there exists an adapted, measurable process (a trading strategy)
$\bar{\psi}$ such that
\[ \bar{\mu} - r \bar{S} = \bar{\sigma}\bar{\sigma}^{\top}\bar{\psi}^{\top} \]
almost everywhere.

$\Box$

\medskip
{\sc Proof of Proposition~\ref{lambdastar-p}}:

Observe that
\[ (\lambda - \bar{\psi}\bar{\sigma})\bar{\sigma}^{\top} =
(\bar{\mu}-r\bar{S})^{\top} - \lambda^{*}\bar{\sigma}^{\top} = 0 \]
almost everywhere.
Hence,
\begin{eqnarray*}
\lambda\lambda^{\top}
& = &
[(\lambda - \bar{\psi}\bar{\sigma}) +\bar{\psi}\bar{\sigma}]
[(\lambda - \bar{\psi}\bar{\sigma}) +\bar{\psi}\bar{\sigma}]^{\top} \\
& = &
(\lambda - \bar{\psi}\bar{\sigma}) (\lambda - \bar{\psi}\bar{\sigma})^{\top} +
\bar{\psi}\bar{\sigma}\bar{\sigma}^{\top}\bar{\psi}^{\top} \\
& \geq &
\bar{\psi}\bar{\sigma}\bar{\sigma}^{\top}\bar{\psi}^{\top} \\
& = &
\lambda^{*}\lambda^{*\top}
\end{eqnarray*}
almost everywhere.

$\Box$

\medskip
{\sc Proof of Theorem~\ref{arfbeta-t}}:

Suppose the trading strategy $\bar{\Delta}$ exists.

Pick a measurable set $\mathcal{N} \subset \Omega \times \mathcal{T}$ with zero measure such that
\[ \bar{\mu} - r\bar{S} = b_{\bar{\Delta}}\bar{\Delta}(\bar{\mu} - r\bar{S}) \]
outside of $\mathcal{N}$.

Suppose $\bar{\gamma}$ is a trading strategy such that on a set $\mathcal{C}$ of positive measure,
$\bar{\gamma}\bar{\sigma}\bar{\sigma}^{\top}\bar{\gamma}^{\top} = 0$ and
$\bar{\gamma}(\bar{\mu}-r\bar{S}) > 0$.
Then $\bar{\mu}-r\bar{S} \neq 0$ and, hence,
$\bar{\Delta}\bar{\sigma}\bar{\sigma}^{\top}\bar{\Delta}^{\top} > 0$ on $\mathcal{C} \setminus \mathcal{N}$.
But then
\[ \bar{\gamma}(\bar{\mu}-r\bar{S}) =
\bar{\gamma} b_{\bar{\Delta}}\bar{\Delta}(\bar{\mu} - r\bar{S}) =
\bar{\gamma}\bar{\sigma}\bar{\sigma}^{\top}\bar{\Delta}^{\top}\frac{1}{
\bar{\Delta}\bar{\sigma}\bar{\sigma}^{\top}\bar{\Delta}^{\top}}
\bar{\Delta}(\bar{\mu} - r\bar{S}) = 0
\]
on $\mathcal{C} \setminus \mathcal{N}$, a contradiction.

Conversely,
if
$(\bar{S},\bar{D})$ is instantaneously arbitrage free,
then it follows from
Proposition~\ref{noinstarbt-p} that
there exists a trading strategy $\bar{\Delta}$ such that
\[ \bar{\mu} - r \bar{S} = \bar{\sigma}\bar{\sigma}^{\top}\bar{\Delta}^{\top} \]
almost everywhere.

Pick a measurable set $\mathcal{N} \subset \Omega \times \mathcal{T}$ with zero measure such that
\[ \bar{\mu} - r \bar{S} = \bar{\sigma}\bar{\sigma}^{\top}\bar{\Delta}^{\top} \]
outside of $\mathcal{N}$.

Set
\[ \mathcal{A} = \{ (\omega,t) \in \Omega \times \mathcal{T}:
\bar{\Delta}(\omega,t)\bar{\sigma}(\omega,t)\bar{\sigma}(\omega,t)^{\top}\bar{\Delta}(\omega,t)^{\top} > 0 \} \]
and
\[ \mathcal{B} = \{(\omega,t) \in \Omega \times \mathcal{T}:
\bar{\Delta}(\omega,t)\bar{\sigma}(\omega,t)\bar{\sigma}(\omega,t)^{\top}\bar{\Delta}(\omega,t)^{\top} = 0 \} \]
Then the indicator functions $1_{\mathcal{A}}$ and $1_{\mathcal{B}}$ are measurable and adapted processes, and in particular, $\mathcal{A}$ and $\mathcal{B}$ are measurable.

On $\mathcal{B} \setminus \mathcal{N}$,
\[ \bar{\mu}-r\bar{S} = \bar{\sigma}\bar{\sigma}^{\top}\bar{\Delta}^{\top} = 0 = b_{\bar{\Delta}}\bar{\Delta}(\bar{\mu}-r\bar{S}) \]
On $\mathcal{A} \setminus \mathcal{N}$,
\[ \bar{\mu}-r\bar{S} = \bar{\sigma}\bar{\sigma}^{\top}\bar{\Delta}^{\top} =
 \frac{\bar{\Delta}(\bar{\mu}-r\bar{S})}{\bar{\Delta}\bar{\sigma}\bar{\sigma}^{\top}\bar{\Delta}^{\top}}\bar{\sigma}\bar{\sigma}^{\top}\bar{\Delta}^{\top} =
b_{\bar{\Delta}}\bar{\Delta}(\bar{\mu}-r\bar{S}) \]
$\Box$

\section{Appendix B: Measurable Linear Algebra}

\label{linalg-app}

A linear equation may have zero, one, or infinitely many solutions.
This appendix shows that where at least one solution exists, a particular solution may be chosen as a measurable function of the parameters of the equation.
In the case where the parameters are stochastic processes, we give a dual characterization of the existence of a solution process.

Let
\[ \mathcal{A}_{M,K} \subset \Re^{M} \times \Re^{M \times K} \]
be the set of pairs $(y,V)$ of an $M$-dimensional column vector $y$ and an $(M \times K)$-dimensional matrix $V$ such that $y$ is in the span of the columns of $V$, or equivalently, such that there exists a $K$-dimensional column vector $x$ with $y = Vx$.

\begin{proposition}
\label{rowsol-p}
The set $\mathcal{A}_{M,K}$ is measurable, and there exists a measurable mapping
\[ \phi_{M,K}: \mathcal{A}_{M,K} \rightarrow \Re^{K} \]
such that $y = V\phi(y,V)$ for all $(y,V) \in \mathcal{A}_{M,K}$.
\end{proposition}

Proposition~\ref{rowsol-p} follows directly from Proposition~\ref{colsol-p} below by transposition.

Let
\[ \mathcal{B}_{M,K} \subset \Re^{K} \times \Re^{M \times K} \]
be the set of pairs $(y,V)$ of a $K$-dimensional row vector $y$ and an $(M \times K)$-dimensional matrix $V$ such that $y$ is in the span of the rows of $V$, or equivalently, such that there exists an $M$-dimensional row vector $x$ with $y = xV$.

\begin{proposition}
\label{colsol-p}
The set $\mathcal{B}_{M,K}$ is measurable, and there exists a measurable mapping
\[ \psi_{M,K}: \mathcal{B}_{M,K} \rightarrow \Re^{M} \]
such that $y = \psi(y,V)V$ for all $(y,V) \in \mathcal{B}_{M,K}$.
\end{proposition}

The proof of Proposition~\ref{colsol-p} will be given below after a series of lemmas.

\begin{proposition}
\label{conj-p}
Let $Y$ and $\Sigma$ be adapted, measurable processes with values in
$\Re^{M}$ and $\Re^{M \times K}$, respectively.
There exists an adapted, measurable process $X$ with values in $\Re^{K}$ such that
$Y = \Sigma X$ almost everywhere if and only if
there does not exist an adapted measurable process $Z$ with values in $\Re^{M}$ such that
$ZY = 1$ and $Z\Sigma = 0$ on a set of positive measure.
\end{proposition}
\begin{proof}
By Proposition~\ref{rowsol-p}, $\mathcal{A}_{M,K}$ is measurable, and the mapping
$\phi_{M,K}: \mathcal{A}_{M,K} \rightarrow \Re^{K}$
is measurable and has the property that $y = V \phi_{M,K}(y,V)$ for all $(y,V) \in \mathcal{A}_{M,K}$.

If $(Y,\Sigma) \in \mathcal{A}_{M,K}$ almost everywhere, then define $X = \phi(Y,\Sigma)$. Then
$X$ is measurable and adapted and $Y = \Sigma X$ almost everywhere. Hence, a process like $X$ exists.

Suppose a process like $X$ exists. Then
almost everywhere, $(Y,\Sigma) \in \mathcal{A}_{M,K}$.
If a process like $Z$ exists, then on a set of positive measure, $(Y,\Sigma) \not\in \mathcal{A}_{M,K}$, a contradiction.
Hence, no process like $Z$ exists.

Finally, if no process like $Z$ exists, then
$(Y,\Sigma) \in \mathcal{A}_{M,K}$ almost everywhere.
To prove this, assume to the contrary that
$(Y,\Sigma) \not\in \mathcal{A}_{M,K}$ on a set of positive measure.

Let $B$ be the set where
$(Y,\Sigma) \not\in \mathcal{A}_{M,K}$.
Then the process $1_{B}$ is measurable and adapted.

By Proposition~\ref{colsol-p},
\[ \mathcal{B}_{M,K+1} \subset \Re^{K+1} \times \Re^{(K+1) \times M} \]
is measurable, and the mapping
\[ \psi_{M,K+1}: \mathcal{B}_{M,K+1} \rightarrow \Re^{M} \]
has the property that
\[ (1,0) = \psi_{M,K+1}((1,0),(y,V))(y,V) \]
for all $(y,V) \in \Re^{M} \times \Re^{M \times K}$ such that
$((1,0),(y,V)) \in \mathcal{B}_{M,K+1}$.
By elementary linear algebra, these are exactly those $(y,V)$ such that $(y,V) \not\in \mathcal{A}_{M,K}$.

Define a process $Z$ by $Z = 0$ outside of $B$ and
\[ Z = \psi_{M,K+1}((1,0),(Y,\Sigma)) \]
on $B$. Since $1_{B}$ is measurable and adapted, so is $Z$. On $B$,
\[ (1,0) = Z(Y,\Sigma) \]
which is equivalent to $ZY = 1$ and $Z\Sigma = 0$.
\end{proof}

\medskip
We proceed to the proof of Proposition~\ref{colsol-p}.

Let $0 \leq r \leq \min\{M,K\}$.

Lemma~\ref{rankh-l} says that we can select, in a measurable manner, a set of $r$ independent linear combinations
of the rows of an $(M \times K)$-dimensional matrices with rank $r$.

Let
$\tilde{D}_{r} \subset \Re^{M \times K}$ denote the set of 
$(M \times K)$-dimensional
matrix with rank $r$.  Then $\tilde{D}_{r}$ is a measurable subset of $\Re^{M \times K}$.

\begin{lemma}
\label{rankh-l}
There exists a measurable mapping
\[ H: \tilde{D}_{r} \rightarrow \Re^{r \times M} \]
such that for each $V \in \tilde{D}_{r}$, the rows of $H(V)V$ span the same linear subspace of $\Re^{K}$ as the rows of $V$.
\end{lemma}
\begin{proof}
Let $j_{1}$ be the first of the numbers $\{1, \ldots ,M\}$ such that
the $j_{1}$th row $V_{j_{1}-}$ of $V$ is non-zero.
Let $j_{2}$ be the first of the numbers $\{j_{1}+1, \ldots ,M\}$ such that
$V_{j_{1}-}$ and $V_{j_{2}-}$ are linearly independent.
Whenever $j_{1}, \ldots ,j_{n}$ have been chosen, and $n < r$,
let $j_{n+1}$ be the first of the numbers $\{j_{n}+1, \ldots ,M\}$ such that
$V_{j_{1}-}$ and $V_{j_{n+1}-}$ are independent.  Set $H(V)_{i} = e^{j_{i}}$ for each $i = 1, \ldots ,r$.
Then $H$ is a measurable mapping, and  for each $i = 1, \ldots ,r$,
\[ H(V)_{i}V = e^{j_{i}}V = V_{j_{i}-} \]
Hence, the $r$ rows of $H(V)V$ are a linearly independent subset of the  rows of $V$.
\end{proof}

\medskip
Lemma~\ref{grammatrix-l}
says that it is possible to orthonormalize the rows of an $(r \times K)$-dimensional matrix with rank $r$ in a measurable manner.

Let
\[ D_{r} \subset \Re^{r \times K} \]
be the set of matrices with rank $r$.

\begin{lemma}
\label{grammatrix-l}
There exists a measurable mapping
\[ G: D_{r} \rightarrow \Re^{r \times r} \]
such that for each $V \in D_{r}$,
the rows of $G(V)V$ are orthonormal
and span the same linear subspace of $\Re^{K}$ as the rows of $V$.
\end{lemma}
\begin{proof}
We shall use the Gramm-Schmidt orthonormalization procedure to
construct $G$ and the mapping
\[ \theta: V \mapsto G(V)V : D_{r} \rightarrow \Re^{r \times K} \]
simultaneously.
The mapping $\theta$ will be measurable and will have the property that
for each $V \in D_{r}$,
the rows of $\theta(V)$ are orthonormal
and have the same span as the rows of $V$.

For notational simplicity, in this proof, write $V_{i} = V_{i-}$ for the $i$th row of $V$, and write $\theta_{i} = \theta(V)_{i-}$ for the $i$th row of $\theta =
\theta(V)$, $i = 1, \ldots ,r$.

Set $x_{1} = V_{1}$ and
\[ \theta_{1} = \frac{1}{\|x_{1}\|}x_{1} \]
Then $\theta_{1}$ is a measurable function of $V$.
Set
\[ G_{1} = \frac{1}{\|x_{1}\|}(1,0, \ldots ,0 ) \]
Then $G_{1}$ is a measurable function of $V$, and
\[ G_{1}V = \frac{1}{\|x\|}V_{1} = \theta_{1} \]

Next, project $V_{2}$ on $\theta_{1}$, let $x_{2}$ be the residual, and let
$\theta_{2}$ be the normalized residual.  Specifically,
\[ V_{2} = t_{2,1}\theta_{1} + x_{2} \]
where $x_{2}$ is orthogonal to $\theta_{1}$.
Now,
\[ V_{2}\theta_{1}^{\top} =
t_{2,1}\theta_{1}\theta_{1}^{\top} + x_{2}\theta_{1}^{\top} =
t_{2,1}\theta_{1}\theta_{1}^{\top} \]
so that
\[ t_{2,1} = 
\frac{V_{2}\theta_{1}^{\top}}{\theta_{1}\theta_{1}^{\top}}
 \]
which is a measurable function of $V$.
Furthermore,
\[ x_{2} = 
 V_{2} - t_{2,1}\theta_{1} \neq 0 \]
since $V_{2}$ and $\theta_{1}$ are independent.
Set
\[ \theta_{2} = \frac{1}{\|x_{2}\|}x_{2} \]
Then
is $\theta_{2}$ is a measurable function of $V$.
Set
\[ G_{2} = \frac{1}{\|x_{2}\|}[(0,1,0,\ldots,0) - t_{2,1}G_{1}] \]
Then $G_{2}$ is a measurable function of $V$, and
\[ G_{2}V = \frac{1}{\|x_{2}\|}[V_{2}-t_{2,1}\theta_{1}] = \frac{1}{\|x_{2}\|}x_{2} = \theta_{2} \]
Once $\theta_{1}, \ldots ,\theta_{n}$ have been chosen, where $n < r$, construct
$\theta_{n+1}$ inductively as follows.
Project $V_{(n+1)}$ on $\theta_{1}, \ldots , \theta_{n}$, let $x_{n+1}$ be the residual,
and let $\theta_{n+1}$ be the normalized residual.  Specifically,
\[ V_{n+1} = \sum_{j=1}^{n}t_{n+1,j}\theta{j} + x_{n+1} \]
where $x_{n+1}$ is orthogonal to $\theta_{j}$ for $j = 1, \ldots ,n$.
Now, for $k = 1, \ldots ,n$,
\[ V_{n+1}\theta_{k} = \sum_{j=1}^{n}t_{n+1,j}\theta{j}\theta_{k} + x_{n+1}\theta_{k} = t_{n+1,k}\theta_{k}\theta_{k}^{\top} \]
so that
\[ t_{n+1,k} = 
\frac{V_{n+1}\theta_{k}^{\top}}{\theta_{k}\theta_{k}^{\top}} \]
which is a measurable function of $V$.  
Furthermore,
\[ x_{n+1} = 
 V_{n+1} - \sum_{j=1}^{n}t_{n+1,j}\theta_{j} \neq 0 \]
since $V_{n+1}$ is not spanned by $\theta_{1}, \ldots ,\theta_{n}$.
Set
\[ \theta_{n+1} = \frac{1}{\|x_{n+1}\|}x_{n+1} \]
Then
is $\theta_{n+1}$ is a measurable function of $V$.
Set
\[ G_{n+1} = \frac{1}{\|x_{n+1}\|}\left[e_{n+1}-\sum_{j-1}^{n}t_{n+1,j}G_{j}\right] \]
Then $G_{n+1}$ is a measurable function of $V$, and
\[ G_{n+1}V =
\frac{1}{\|x_{n+1}\|}\left[V_{n+1}-\sum_{j-1}^{n}t_{n+1,j}\theta_{j}\right] =
\frac{1}{\|x_{n+1}\|}x_{n+1} = \theta_{n+1} \]
This completes the construction of $G$ and $\theta$.
\end{proof}

\medskip
{\sc Proof of Proposition~\ref{colsol-p}}:

For each $r$ with $0 \leq r \leq \min\{M,K\}$, let
\[ \mathcal{B}_{r} = \{ (y,V) \in \mathcal{B}_{M,K}: \mbox{rank}(V) = r \} \]
Then
$\mathcal{B}_{r}$ is measurable, and
$\mathcal{B}_{M,K}$ is measurable since
\[ \mathcal{B}_{M,K} = \bigcup_{r = 0}^{\min\{M,K\}}\mathcal{B}_{r} \]

To complete the proof, it suffices to show that for each
$r$ with $0 \leq r \leq \min\{M,K\}$, there exists
a measurable mapping
\[ \psi_{r}: \mathcal{B}_{r} \rightarrow \Re^{M} \]
such that $y = \psi_{r}(y,V)V $ for all $(y,V) \in \mathcal{B}_{r}$.

Let
\[ H: \tilde{D}_{r} \rightarrow \Re^{r \times M} \]
be the mapping from Lemma~\ref{rankh-l}, and let
\[ G: D_{r} \rightarrow \Re^{r \times r} \]
be the mapping from Lemma~\ref{grammatrix-l}.
Then the mapping
\[ J: \tilde{D}_{r} \rightarrow \Re^{r \times M}: V \mapsto G(H(V)V)H(V) \]
is measurable and has the property that for each $V \in \tilde{D}_{r}$, the $r$ rows of $J(V)V$ are orthonormal, or equivalently, $J(V)V V^{\top} J(V)^{\top} = I_{r}$, the $(r \times r)$-dimensional unit matrix.
Furthermore, the rows of $J(V)V$ span the same linear subspace of $\Re^{K}$ as the rows of $V$.  Hence, if $(y,V) \in \mathcal{B}_{r}$, then there exists $z \in \Re^{r}$ such that $y = zJ(V)V$.
This implies that
\[ y V^{\top}J(V)^{\top} = zJ(V)VV^{\top}J(V)^{\top} = z \]
and
\[ y = zJ(V)V = y V^{\top}J(V)^{\top}J(V)V \]
Define the mapping
\[ \psi_{r}: \mathcal{B}_{r} \rightarrow \Re^{M} \]
by
\[ \psi_{r}(y,V) = 
y V^{\top} J(V)^{\top}J(V) \]
Then $\psi_{r}$ is measurable, and for all $(y,V) \in \mathcal{B}_{r}$,
\[ \psi_{r}(y,V)V = y V^{\top} J(V)^{\top}J(V)V = y \]
$\Box$

\bibliographystyle{plain}
\bibliography{bookrefs,bookrefsadd,consolidated}

\end{document}